\newcommand{\be}{\begin{equation}}\newcommand{\ee}{\end{equation}}
\newcommand{\bea}{\begin{eqnarray}}\newcommand{\eea}{\end{eqnarray}}
\newcommand{\nn}{\nonumber}\newcommand{\p}[1]{(\ref{#1})}
\documentstyle[12pt]{article}

\begin{document}
\renewcommand{\thefootnote}{\fnsymbol{footnote}}
\thispagestyle{empty}
\begin{center}
International Atomic Energy Agency \\
and \\
United Nations Educational Scientific and Cultural Organization \\
INTERNATIONAL CENTRE FOR THEORETICAL PHYSICS \vspace{1.5cm} \\
{\hfill  hep-th/9406005   }\vspace{1cm} \\
{\bf LINEARIZING  $W$-ALGEBRAS} \vspace{1.5cm} \\
S.O. Krivonos\footnote{E-mail: krivonos@thsun1.jinr.dubna.su} \\
Bogoliubov Laboratory of Theoretical Physics, JINR, Dubna, Russia
\vspace{1cm}\\
and \vspace{1cm} \\
A.S. Sorin\footnote{E-mail: sorin@thsun1.jinr.dubna.su}{}\footnote{Permanent
address: Bogoliubov Laboratory of Theoretical Physics, JINR, Dubna, Russia.}\\
International Centre for Theoretical Physics, Trieste, Italy \vspace{2cm}\\
{\bf Abstract}
\end{center}
We show that the Zamolodchikov's and Polyakov-Bershadsky
nonlinear algebras $W_3$ and $W_3^{(2)}$ can be embedded as
subalgebras into some {\em linear} algebras with finite set of currents.
Using these linear algebras  we find new field realizations of $W_3^{(2)}$
and $W_3$ which could be a starting point for constructing new versions
of $W$-string theories. We also reveal a number of hidden
relationships between $W_3$ and $W_3^{(2)}$. We conjecture that
similar linear algebras can exist for other $W$- algebras
as well. \vspace{1cm}\\
\begin{center}
{\it Submitted to Phys. Lett. B} \\
\vfill
MIRAMARE-TRIESTE \\
June 1994
\end{center}
\setcounter{page}0
\renewcommand{\thefootnote}{\arabic{footnote}}
\setcounter{footnote}0
\newpage

\section{Introduction}

By now there exist many examples of extended conformal algebras, the
overwhelming majority of which (the so called $W$-type algebras)
is essentially
nonlinear (see e.g. \cite{BS} and references therein).
Because of the intrinsic
nonlinearity of W-algebras, their study is a  more difficult
task compared to linear algebras.
However, as was shown by Goddard and Schwimmer \cite{GS},
there exists a deep relation between  $SO(3)$ and
$SO(4)$ extended nonlinear superconformal algebras of
Knizhnik-Bershadsky \cite{KB} on the one hand and
linear $N=3$ and $N=4$ superconformal algebras \cite{Ademollo} on the other:
the latter can be obtained from the nonlinear
algebras by adding some spin $1/2$ and spin 0 currents.
So, there arises a natural question: whether in other cases there exist
linear Lie algebras with finite number of currents,
such that they contain some nonlinear W-algebras with an arbitrary
central charge as subalgebras in a nonlinear basis.
If existing, such linear algebras could be helpful at least for
constructing new realizations of $W$-algebras.
In the present letter we give the positive answer
to the above question for the two
simplest nonlinear algebras: Zamolodchikov $W_3$ algebra \cite{W3}
and Polyakov-Bershadsky $W_3^{(2)}$ algebra \cite{W32}.
We find that in both these cases the linear algebras indeed exist
and contain the considered $W$-algebras as subalgebras.
Using the  linear algebras obtained we construct new field realizations of
$W_3^{(2)}$ and $W_3$ algebras and find a number of intrinsic relationships
between these algebras.

\setcounter{equation}0
\section{Linearizing $W_3$ algebra.}

In this Section we will construct a linear algebra $W_3^{lin}$ which contain
$W_3$ algebra as a subalgebra in some basis for the defining currents.
Hereafter, some nonlinear redefinition
of the currents is called the change of the basis of the (non)linear algebra,
if (i) it is invertible and (ii) both it and its inverse are polynomial
in the currents and derivatives of the latter. A subset of the currents is
meant to form a (non)linear subalgebra of  given $W$-algebra if in some
basis this subset is closed; all the algebras related by (nonlinear)
transformations of the basis are treated as equivalent.

The $W_3$ algebra \cite{W3} contains the currents $\left\{ T , W \right\}$
with  spins $\left\{ 2,3\right\}$ respectively, obeying the following
operator product expansions (OPE):\footnote{All the currents in the
right-hand sides of the OPEs are evaluated at the point
$z_2$, $z_{12}=z_1-z_2$. Multiple composite operators (AB) are always
ordered from the right to the left.}
\begin{eqnarray}
T(z_1)T(z_2) & = &  \frac{c}{2}\frac{1}{z_{12}^4}+
        \frac{2T}{z_{12}^2}+\frac{T'}{z_{12}}  \quad , \quad
T(z_1)W(z_2)  =   \frac{3W}{z_{12}^2}+\frac{W'}{z_{12}} \;, \nn \\
W(z_1)W(z_2) & = &  \frac{c}{3}\frac{1}{z_{12}^6}+
    \frac{2T}{z_{12}^4}+\frac{T'}{z_{12}^3}+
    \left[2\kappa\Lambda+\frac{3}{10}T''\right]\frac{1}{z_{12}^2}+
    \left[\kappa\Lambda'+\frac{1}{15}T'''\right]\frac{1}{z_{12}} \;,
     \label{W3}
\end{eqnarray}
where
\begin{equation}
 \Lambda \equiv (T\;T)-\frac{3}{10}T'' \quad , \quad
 \kappa \equiv \frac{16}{22+5c} \quad .
\end{equation}

Our main  idea is to extend a nonlinear algebra by  some currents
with {\em a priori} unknown properties to get a linear algebra after
passing to another basis for the currents. The only assumption we will
impose from the beginning is that the nonlinear algebra we started with
still forms a subalgebra (in the above sense) in this extended algebra.

Fortunately, for the case of $W_3$ algebra it proves enough to add only one
extra spin 1 current $J$ with the following OPE's:
\begin{eqnarray}
T(z_1)J(z_2) & = & \frac{1+3c_1}{z_{12}^3}+
     \frac{J}{z_{12}^2}+\frac{J'}{z_{12}} \quad , \quad
J(z_1)J(z_2)  =   \frac{c_1+1}{z_{12}^2} \;, \nn \\
W(z_1)J(z_2) & = & \sqrt{\frac{2}{3-22c_1-45c_1^2}}\left\{
       -\frac{(1+3c_1)^2}{2}\frac{1}{z_{12}^4} -\frac{(1+3c_1)J}{z_{12}^3}
       +\left( 2(1+c_1)T -2(JJ)+ \right. \right. \nn \\
 & & \left. \frac{1+3c_1}{2}J'\right)\frac{1}{z_{12}^2} -
  \left(  \sqrt{\frac{3-22c_1-45c_1^2}{2}}W - 2 (TJ) -
   \frac{3+c_1}{2}T'+ \right. \nn \\
  & & \left.\left.  \frac{4}{3(1+c_1)}(JJJ) -
  \frac{2(c_1-1)}{1+c_1} (J'J)-
     \frac{3c_1-1}{3(1+c_1)}J''\right)
   \frac{1}{z_{12}}\right\}  \quad , \label{W3l}
\end{eqnarray}
where the central charges $c$ and $c_1$ are connected as
\begin{equation}
c=2-\frac{4(3c_1+1)^2}{c_1+1} \quad . \label{ccr}
\end{equation}
Let us stress that the Jacobi identities completely fix all coefficients
in the OPE's \p{W3l} as well as the relation between central charges \p{ccr}.
Thus, there is
a unique possibility to extend the $W_3$ algebra by the spin 1 current
under the natural assumption of preserving the structure of $W_3$.

Now it is a matter of straightforward calculation to show that this new
algebra \p{W3},\p{W3l} which we denote as $W_3^{lin}$
is actually a linear algebra. To this end,
we do the following {\em invertible} nonlinear transformation to the
new basis $\left\{ J,T_0,G^{+} \right\}$:\footnote{Let us note that
transformation from $T$ to $T_0$ is linear, but current $J$ becomes
primary with respect to $T_0$. This is the reason why we prefer
to deal with $T_0$ rather than with $T$.}
\begin{eqnarray}
T_0 & = & T +\frac{3c_1+1}{2(c_1+1)} J' \;, \nn \\
G^{+} & = & W - \frac{2\sqrt{2}}{(8(c_1+1)-5(3c_1+1)^2)^{1/2}}
        \left\{ (JT_0)-\frac{2}{3(c_1+1)}(JJJ)
         +\frac{3c_1+1}{2(c_1+1)}(JJ') - \right. \nn \\
  & & \left. \frac{3c_1+1}{4}\left( T_0+\frac{3c_1+1}{6(c_1+1)}
     J'\right)' \right\}. \label{realiz}
\end{eqnarray}
In this new basis  OPE's \p{W3},\p{W3l} become linear
\begin{eqnarray}
T_0(z_1)T_0(z_2) & = &
 \frac{1-4c_1-9c_1^2}{2(c_1+1)}\frac{1}{z_{12}^4}+
        \frac{2T_0}{z_{12}^2}+\frac{T_0'}{z_{12}} \quad , \quad
T_0(z_1)G^{+}(z_2)  =
  \left[ \frac{3}{2}+\frac{1}{c_1+1} \right]\frac{G^{+}}{z_{12}^2}+
           \frac{G^{+}{}'}{z_{12}} \;, \nn \\
T_0(z_1)J(z_2) & = &
       \frac{J}{z_{12}^2}+\frac{J'}{z_{12}} \quad , \quad
J(z_1)J(z_2)  =   \frac{c_1+1}{z_{12}^2}  \quad , \quad
J(z_1)G^{+}(z_2)  =  \frac{G^{+}}{z_{12}} \label{lin}.
\end{eqnarray}

So we have explicitly shown that $W_3^{lin}$ algebra (2.1)-(2.4) is linear
and contains the $W_3$ algebra as a subalgebra in the nonlinear basis.

We close this Section with a few  comments.

First of all, let us note that the linear $W_3^{lin}$ algebra \p{lin} is
homogeneous in the current $G^{+}$. So $G^{+}$ is the typical null field
and we can consider  the limit $G^{+} = 0$. In this limit
the expressions \p{realiz} give the well known  realization of $W_3$ algebra
in terms of the spin 1 current $J$ and an arbitrary stress-tensor
$T_0$ \cite{Rom}.

Secondly, we would like to stress that, strictly speaking, it make sense
to fix the  conformal spins of the currents only in the basis where all of
them are at least quasi-primary. So, while we consider the $W_3$
algebra \p{W3}, the current $W$ has spin 3, but in the case of
extended algebra \p{W3},\p{W3l}, where the current $J$ is not quasi-primary,
we must fix all spins with respect to the new stress-tensor
$T_0$ \p{realiz}. In this basis the conformal spin of current $W$
(as well as  of the current $G^{+}$ related to $W$ by the invertible
transformation \p{realiz}) is just\footnote{It becomes 3/2 in the classical
limit $c_1 \rightarrow \infty$.}
$$\frac{3}{2}+\frac{1}{c_1+1} \;.$$
We have explicitly checked that there exist no other non-trivial
extensions of $W_3$ algebra by a primary spin 1 current,
such that the structure of $W_3$ itself remains intact.

Finally, we can reverse our arguments.
We may start from the linear algebra \p{lin} and show that
the currents $T$ and $W$ defined through the
transformations \p{realiz} form the $W_3$ algebra \p{W3}.
Thus, all the remarkable nonlinear features of $W_3$ algebra can be
traced to the choice of a nonlinear basis in the linear algebra \p{lin}.

In the next Section we will use this approach to construct the linear
algebra $W_3^{(2)lin}$ which contains quantum $W_3^{(2)}$ algebra
as a subalgebra.

\setcounter{equation}0
\section{Linearizing $W_3^{(2)}$ algebra.}

In this Section we explicitly demonstrate that the quantum $W_3^{(2)}$
algebra is a subalgebra of very simple {\em linear} algebra.

The algebra $W_3^{(2)}$ contains the bosonic currents
$\left\{J_w, G^{+}, G^{-}, T_w\right\}$ with the spins
$\left\{1, 3/2, 3/2, 2\right\}$,
respectively \cite{W32}, and supplies the simplest nontrivial example of
algebras with fractional bosonic spins.
The quantum OPE's for its currents are as follows
\begin{eqnarray}
T_w(z_1)T_w(z_2) & = & \left[ \frac{(7-9c_1)c_1}{2(c_1+1)}\right]
       \frac{1}{z_{12}^4}+
        \frac{2T_w}{z_{12}^2}+\frac{T_w'}{z_{12}} \quad , \quad
T_w(z_1)G^{\pm}(z_2)  =   \frac{3}{2}\frac{G^{\pm}}{z_{12}^2}+
           \frac{G^{\pm}{}'}{z_{12}} \, \nn \\
T_w(z_1)J_w(z_2) & = &  \frac{J_w}{z_{12}^2}+\frac{J_w'}{z_{12}} \; , \;
J_w(z_1)J_w(z_2)  =   \frac{c_1}{z_{12}^2}  \; , \;
J_w(z_1)G^{\pm}(z_2)  =  \pm \frac{G^{\pm}}{z_{12}} \\
G^{+}(z_1)G^{-}(z_2) & = & \left[ \frac{(3c_1-1)c_1}{c_1+1}\right]
     \frac{1}{z_{12}^3}+
       \left[ \frac{3c_1-1}{c_1+1}\right] \frac{J_w}{z_{12}^2}- \nn \\
            & & \left[ T_w-
      \frac{2}{c_1+1}(J_wJ_w)
       -\frac{3c_1-1}{2(c_1+1)}J_w'\right]\frac{1}{z_{12}}. \label{w32}
\end{eqnarray}
Without going into details, let us show that the linear algebra, which we will
denote as $W_3^{(2)lin}$ and which is defined by the following set of OPE's
\begin{eqnarray}
T_w(z_1)T_w(z_2) & = & \left[ \frac{(7-9c_1)c_1}{2(c_1+1)}\right]
     \frac{1}{z_{12}^4}+\frac{2T_w}{z_{12}^2}+\frac{T_w'}{z_{12}}
  \quad , \quad
T_w(z_1)J_w(z_2)  =   \frac{J_w}{z_{12}^2}+\frac{J_w'}{z_{12}} \; , \nn \\
T_w(z_1)\tilde{G}^{+}(z_2) & = &  \frac{3}{2}\frac{\tilde{G}^{+}}{z_{12}^2}+
           \frac{\tilde{G}^{+}{}'}{z_{12}} \quad , \quad
T_w(z_1)G^{-}(z_2)  =   \frac{3}{2}\frac{G^{-}}{z_{12}^2}+
           \frac{G^{-}{}'}{z_{12}} \; , \nn \\
J_w(z_1)J_w(z_2) & = &  \frac{c_1}{z_{12}^2}  \quad , \quad
J_w(z_1)\tilde{G}^{+}(z_2)  =   \frac{\tilde{G}^{+}}{z_{12}} \quad , \quad
J_w(z_1)G^{-}(z_2)  =  - \frac{G^{-}}{z_{12}} \; , \nn \\
\tilde{G}^{+}(z_1)G^{-}(z_2) & = & \mbox{regular}\; , \label{l1} \\
T_w(z_1)\gamma(z_2) & = &  -\frac{1}{2}\frac{\gamma}{z_{12}^2}+
           \frac{\gamma'}{z_{12}} \quad , \quad
J_w(z_1)\gamma(z_2)  =   \frac{\gamma}{z_{12}} \; , \label{l2} \\
G^{-}(z_1)\gamma(z_2) & = &  \frac{1}{z_{12}}   \label{l3}
\end{eqnarray}
for the bosonic currents
$\left\{\gamma, J_w, G^{-}, \tilde{G}^{+}, T_w \right\}$
with the spins $\left\{-1/2, 1, 3/2, 3/2, 2 \right\}$, respectively, contains
$W_3^{(2)}$ algebra (3.1),\p{w32} as a subalgebra.

In order to prove
this, we do the following {\em invertible} nonlinear transformation to the
new basis $\left\{\gamma, J_w, G^{-}, G^{+}, T_w\right\}$, where
\begin{eqnarray}
G^{+} & = & \tilde{G}^{+}+ (T_w\gamma )-\frac{2}{c_1+1}(J_wJ_w\gamma )-
       \frac{3c_1+7}{2(1+c_1)}(J_w'\gamma )+
     \frac{2}{1+c_1}(J_wG^{-}\gamma\gamma) - \nn \\
     & &  \frac{2}{3(c_1+1)}(G^{-}G^{-}\gamma\gamma\gamma) -
       3(G^{-}\gamma'\gamma )+ \frac{1-c_1}{1+c_1}(G^{-}{}'\gamma\gamma )
          +3\left((J_w\gamma )-\frac{c_1+1}{2}\gamma'\right)' .
\end{eqnarray}
In this basis the complete set of OPE's of $W_3^{(2)lin}$ algebra
is given by eqs. (3.1),\p{w32},\p{l2},\p{l3} and by the following OPE
\begin{equation}
G^{+}(z_1)\gamma(z_2)=-\frac{(\gamma\gamma )}{z_{12}^2}+
  \left[ \frac{2}{3(1+c_1)}(G^{-}\gamma\gamma\gamma )-
    \frac{2}{1+c_1}(J_w\gamma\gamma )+
       (\gamma'\gamma )\right] \frac{1}{z_{12}} \; .
\end{equation}
So one observes that $W_3^{(2)lin}$ indeed contains
$W_3^{(2)}$ as a subalgebra.

Let us remark that two currents $G^{-}$ and $\gamma$ looks like
``ghost--anti-ghost'' fields and so $W_3^{(2)lin}$  algebra
can be simplified by means of the standard ghost decoupling
transformations
\begin{eqnarray}
J & = & J_w - (G^{-}\gamma ) \quad , \nn \\
T_0 & = & T_w -\frac{3}{2}(G^{-}\gamma' )-\frac{1}{2}(G^{-}{}'\gamma )-
         \frac{1}{c_1+1}\left( J_w-(G^{-}\gamma )\right)' \; .
\end{eqnarray}
After this $W_3^{(2)lin}$ algebra splits into  the direct product
of the bosonic ghost--anti-ghost $\left\{\gamma, G^{-}\right\}$
algebra \p{l3} and the algebra of the
currents $\left\{ J, T_0, G^{+}\right\}$ which coincides with
$W_3^{lin}$ algebra \p{lin}.
So
\begin{equation}
W_3^{(2)lin}=\Gamma  \otimes W_3^{lin} \quad , \quad
  \Gamma=\left\{\gamma, G^{-}\right\} \; , \;
   W_3^{lin}= \left\{J,T_0, G^{+}\right\} \; .\label{ww}
\end{equation}
Thus, $W_3$  is also a subalgebra of $W_3^{(2)lin}$ since
$W_3^{lin}\subset W_3^{(2)lin}$. In this sense
$W_3^{(2)lin}$ is analogous to the affine $sl(3)$ algebra, which can be
reduced via the DS hamiltonian reduction procedure
to either $W_3$ or $W_3^{(2)}$. However, unlike it, $W_3^{(2)lin}$
contains $W_3$ and $W_3^{(2)}$ as subalgebras. This means that every
realization of $W_3^{(2)lin}$ is a realization of the $W_3$ and $W_3^{(2)}$
algebras simultaneously
\footnote{Of course, the inverse statement is not correct in general.}.
So the problem of the construction of realizations of
these subalgebras is reduced to the problem of constructing  realizations
of $W_3^{(2)lin}$. Owing to the very simple structure
of $W_3^{(2)lin}$ \p{ww}, this
task actually amounts to the construction of realizations of $W_3^{lin}$.
In the rest of this Section we will present an example of such a
realization.

    From the simple structure of the $W_3^{lin}$ algebra OPEs \p{lin} it is
evident that its most general realization includes at least
two free bosonic scalar
fields $\phi_i$ ($i=1, 2$) with OPEs
\begin{equation}
\phi_i(z_1)\phi_j(z_2) = -\delta_{ij}ln(z) \; ,
\end{equation}
as well as a commuting with them Virasoro stress tensor $\tilde{T}$ having a
nonzero
central charge which we denote $c_T$.
Representing  the bosonic primary field $G^{+}$
in the standard way by an exponential of $\phi_i$,
$J$ by the derivatives of $\phi_i$ and $T_0$ by the sum
of $\tilde{T}$ and
the standard stress-tensors of $\phi_i$ with background charges, and requiring
them to satisfy the OPEs \p{lin}, we find the following expressions
\begin{eqnarray}
G^{+} & = &s \cdot \mbox{exp}\left(i\sqrt{N-\frac{1}{c_1+1}}\phi_2+
            \frac{i}{\sqrt{c_1+1}}\phi_1\right) , N \in Z \quad ,\nn \\
J & = &i\sqrt{c_1+1}\phi_1'\quad , \nn \\
T_0 & = & \tilde{T}-\frac{1}{2}\left( \phi_1'\right)^2  -
      -\frac{1}{2}\left( \phi_2'\right)^2  -
 \frac{i\left(3-N+\frac{2}{c_1+1}\right)}{2\sqrt{N-\frac{1}{c_1+1}}}\phi_2''
             \quad , \nn \\
c_T & = & 3\frac{\left( 3-N+\frac{2}{c_1+1}\right)^2}{N-\frac{1}{c_1+1}}-
      \frac{(3c_1+1)^2}{c_1+1} \quad , \label{ff}
\end{eqnarray}
where $s$ is an arbitrary parameter. Its arbitrariness reflects the invariance
of the OPEs \p{lin} with respect to rescaling of the null field $G^{+}$.
If $s\neq 0$, it can always be chosen, e. g., equal to unity by a constant
shift
of the field $\phi_1$.

   In the case of $s=0$ the obtained realization can be simplified
by introducing a new Virasoro stress-tensor $T_n$ which absorbs the field
$\phi_2$
\begin{eqnarray}
T_n & = & \tilde{T}-\frac{1}{2}\left( \phi_2'\right)^2  -
 \frac{i\left(3-N+\frac{2}{c_1+1}\right)}{2\sqrt{N-\frac{1}{c_1+1}}}\phi_2''
     \quad , \\
c_{T_n} & = & 1 - \frac{(3c_1+1)^2}{c_1+1} \label{cc}
\end{eqnarray}
and $\phi_2-$dependence disappears altogether. In this notation the
expressions \p{ff} are given by
\begin{eqnarray}
G^{+} & = & 0 \quad , \nn \\
J & = & i\sqrt{c_1+1}\phi_1' \quad , \nn \\
T_0 & = & T_n -\frac{1}{2}\left( \phi_1' \right)^2 \quad . \label{ff1}
\end{eqnarray}

After substituting eqs. \p{ff} into \p{realiz}, we get a realization of the
$W_3$ algebra which generalizes the realization obtained in \cite{Rom} and is
reduced to it at $s=0$.
After substituting \p{ff} into (3.6),(3.8) we get a new realization of
the $W_3^{(2)}$ algebra. The realizations constructed here
may be important in the $W_3$ ($W_3^{(2)}$)-string theories, but detailed
consideration of them is beyond the scope of the present letter.

We close this Section with several comments.

First, in the case of $G^{+}=0$, the
algebra $W_3^{lin}$  \p{lin} reduces to a direct product of the $u(1)$
algebra and the Virasoro one with the central charge \p{cc}.
Then, the minimal Virasoro models \cite{min} which correspond to
the central charge
\begin{equation}
c_m=1-6\frac{(p-q)^2}{pq} \Rightarrow c_1=\frac{2}{3}\frac{q}{p}-1
\end{equation}
give rise to the following induced central charges
\begin{eqnarray}
c_{W_3} & = & 2\left( 1 - 12 \frac{(p-q)^2}{pq} \right)  \label{cc1}\\
c_{W_{3/2}} & = &  \left( 1 - 6 \frac{(q-2p)^2}{pq} \right) \label{cc2}
\end{eqnarray}
for the $W_3$ and $W_3^{(2)}$ subalgebras of $W_3^{(2)lin}$, respectively.
The expression \p{cc1} exactly
coincides with the central charges of $W_3$ minimal
models \cite{min1}, eq. \p{cc2} reproduces
the minimal $W_3^{(2)}$ models \cite{min2} at $q=2\tilde{q}$.

Secondly, it is easy to obtain the classical analogues of all
quantum expressions constructed here. We will not write all
these relations except for the OPE's between the
currents of classical version of the algebra $W_3^{lin}$ \p{lin} :
\begin{eqnarray}
T_0(z_1)T_0(z_2) & = &  \frac{-\frac{9}{2}c_1}{z_{12}^4}+
        \frac{2T_0}{z_{12}^2}+\frac{T_0'}{z_{12}}  \quad , \quad
T_0(z_1)J(z_2)  =   \frac{J}{z_{12}^2}+\frac{J'}{z_{12}} \; ,\nn \\
T_0(z_1)G^{+}(z_2) & = &  \frac{3}{2}\frac{G^{+}}{z_{12}^2}+
           \frac{G^{+}{}'}{z_{12}} , \;
J(z_1)J(z_2)  =   \frac{c_1}{z_{12}^2} , \;
J(z_1)G^{+}(z_2)  =  \frac{G^{+}}{z_{12}}.  \label{cl1}
\end{eqnarray}
It is interesting to note that these OPE's form a subalgebra of
the classical $W_3^{(2)}$ following from (3.1)-(3.2) in the limit
$c_1 \rightarrow \infty$. So, in the classical case
both $W_3^{lin}$ and its nonlinear subalgebra $W_3$ are
subalgebras of  $W_3^{(2)}$. This statement fails to be true
in the quantum case. Only a weaker property $W_3^{lin}\subset W_3^{(2)lin}$
is retained.

\setcounter{equation}0
\section{Conclusions and outlook.}

   In this paper we have constructed the linear conformal algebras
with finite set of currents  which contain
the  $W_3^{(2)}$ and $W_3$ algebras as subalgebras in some nonlinear basis.
We have also found new realizations of $W_3^{(2)}$ and $W_3$  which can
be a starting point  for  construction of new versions
of $W$-string theories. The study of these linear algebras allowed us
to reveal hidden intrinsic relations between them as well as between
$W_3^{(2)}$ and $W_3$ algebras.
The linear algebras constructed are
similar in many aspects to the affine Kac-Moody algebra $sl(3)$ which
is reducible both to the  $W_3^{(2)}$ and $W_3$ algebras.
However, the essential difference consists in that
the linear algebras include $W_3^{(2)}$ and $W_3$  as subalgebras.
We believe that the most of properties of nonlinear algebras
and the theories constructed on their basis, could be understood in a
simpler way by studying their linear counterparts.

We do not still know whether the  existence of such linear algebras
is a general property of
the $W$-type algebras or an artifact of some exceptional cases like
$W_3$ and $W_3^{(2)}$. If this property is general, there must exist
an algorithmic procedure for constructing  the linear algebras.
In any case, it seems important to have more examples of linearization
as well as real applications of these linear algebras.
In the forthcoming paper we will consider  more examples of nonlinear
algebras admitting a linearization.\vspace{1cm}

\indent{\large\bf Acknowledgments.}

During the course of this work we had many discussions which are
greatly appreciated. We thank  S.Bellucci, F.Delduc, V.Fateev,
L.Feher, J.-L.Gervais, A.Honecker, A.Isaev, E.Ivanov, P.Kulish, J.Lukierski,
W.Nahm,
V.Ogievetsky, M.Olshanetsky, A.Pashnev, V.Rittenberg, M.Va\-s\-i\-liev,
D.Volkov, G.Zinovjev for many useful conversations.

It is a pleasure for us to especially thank E.Ivanov for careful reading
of the manuscript and numerous comments which helped us to make the
paper more readable.

This work was partially supported by grant 93-02-03821 of RFFR.
We are also indebted to the Laboratori Nazionali di Frascati for hospitality
and partial financial support extended to us during the final part
of this work.

One of the authors (A.S.S.) would like to thank Professor Abdus Salam,
the International Atomic Energy Agency and UNESCO for hospitality at
the International Centre for Theoretical Physics, Trieste.

\end{document}